\documentclass[pre,aps,epsfig,byrevtex,amsmath,amssymb,preprint]{revtex4}
\usepackage{times}
\usepackage{epsf}
\usepackage{graphicx}
\usepackage{dcolumn}
\usepackage{bm}

\begin{document}

\date {\today}

\title{Theory of second harmonic generation in colloidal crystals}
\author{J. P.
Huang\footnote{Corresponding author. Electronic address:
jphuang@fudan.edu.cn}, Y. C. Jian, C. Z. Fan}
\address{Surface Physics Laboratory  and Department of Physics, Fudan University, Shanghai
200433, China}
\author{K. W. Yu}
\address{Department of Physics and Institute of Theoretical Physics, The Chinese University of Hong Kong,
Shatin, New Territories, Hong Kong}

\begin {abstract}
On the basis of the Edward-Kornfeld formulation, we study the
effective susceptibility of second-harmonic generation (SHG) in
colloidal crystals, which are made of graded metallodielectric
nanoparticles with an intrinsic SHG susceptibility suspended in a
host liquid. We find a large enhancement and redshift of SHG
responses, which arises from the periodic structure, local field
effects and gradation in the metallic cores. The optimization of the
Ewald-Kornfeld formulation is also investigated.
\end{abstract}

\maketitle

\section{Introduction}\label{intro}

Nonlinear composite materials with high nonlinear optical
susceptibilities or optimal figure of merit (FOM) have drawn
considerable attention for their potential applications, e.g., in
bistable switches, optical correlators, and so on$^{1-6}$. With
several advancements in nanotechnology, such as templated
sedimentation and dielectrophoresis$^{7-10}$, it is possible to
fabricate particles with specific geometry. It has also been
reported$^{4,6,11,12}$ that graded composite materials whose
physical properties vary gradually in space$^{13-16}$ can exhibit
enhanced nonlinear optical responses$^6$ and optimal dielectric ac
responses$^{17}$, as well as conductive responses$^{18}$.

It is known that materials lacking inversion symmetry can exhibit
a so-called second order nonlinearity$^1$. This can give rise to
the phenomenon of second harmonic generation (SHG), i.e., an input
(pump) wave can generate another wave with twice the optical
frequency (namely, half the wavelength) in the medium. In most
cases, the pump wave is delivered in the form of a laser beam, and
the second-harmonic wave is generated in the form of a beam
propagating in a similar direction. The physical mechanism behind
SHG can be understood as follows. Due to the second order
nonlinearity, the fundamental (pump) wave generates a nonlinear
polarization which oscillates with twice the fundamental
frequency. According to Maxwell's equations, this nonlinear
polarization radiates an electromagnetic field with this doubled
frequency. Due to phase matching issues, the generated
second-harmonic field propagates dominantly in the direction of
the nonlinear polarization wave. The latter also interacts with
the fundamental wave, so that the pump wave can be attenuated
(pump depletion) when the second-harmonic intensity develops. In
the mean time, the energy is transferred from the pump wave to the
second-harmonic wave.  The SHG effect, like the third-order
Kerr-type coefficient, involves the nonlinear susceptibilities of
the constituents and local field enhancement which arises from the
structure of composite materials$^{19-23}$. For example, Hui and
Stroud studied a dilute suspension of coated particles with the
shell having a nonlinear susceptibility for SHG$^{19}$, and Fan
and Huang designed a class of ferrofluid-based soft nonlinear
optical materials with enhanced SHG with magnetic-field
controllabilities$^{20}$. Both SHG and phase transformation
behaviors can be detected in many nanocrystals by using  classical
methods like X-ray diffraction or Raman spectroscopy$^{24,25}$.
Until now, achieving enhanced SHG is a challenge$^{26,27}$.

Theoretical$^{28}$ and experimental$^{29,30}$ reports also
suggested that spherical particles exhibit a rather unexpected and
nontrivial SHG due to the broken inversion symmetry at particle
surfaces (namely, atoms in the surface occupy positions that lack
inversion symmetry), despite their central symmetry which
seemingly prohibits second-order nonlinear effects. In colloidal
suspensions, the SHG response for centrosymmetric particles was
experimentally reported$^{29,30}$. Most recently, SHG was also
shown to appear for spherical semiconductor nanocrystals$^{31}$.
So far, the SHG arising from centrosymmetrical structure has
received an extensive attention$^{28-30,32,33}$.




Colloidal crystals have been widely studied in nanomaterials and
have potential applications in nanophotonics, chemistry, and
biomedicine$^{34}$. For colloidal crystals, the individual
colloidal nanoparticles should be touching since the lattice
parameters of the crystals, $a$, $b$ and $c$, should satisfy the
geometric constraint $a^2+b^2+c^2 = 16R_0^2$, where $R_0$ denotes
the radius of an individual colloidal nanoparticle. However, it is
possible to achieve a colloidal crystal without the particles'
touching if the colloidal nanoparticles are charged and stabilized
by electrostatic forces. In this work, we shall investigate
colloidal crystals with the particles' touching. Owing to recent
advancements in the fabrication of nanoshells$^{35,36}$, we are
allowed to use a dielectric surface layer with thickness $d$ on a
graded metallic core with radius $a_0$, in order to activate
repulsive or attractive forces between the nanoparticles. The
dielectric constant of the metallic core should be a radial
function because of a radial gradation, and that of the surface
layer can be the same as that of the host liquid, as to be used in
this work. The latter is also a crucial requirement because
otherwise multipolar interaction between the metallic cores can
become important$^{37}$. In this regard, the surface layer
contributes to the geometric constraint, rather than the effective
optical responses of the colloidal crystals.
In this work, based on the Ewald-Kornfeld formulation
[Eq.~(\ref{Ewald1})]$^{38}$ and the nonlinear differential
effective dipole approximation (NDEDA)
 method [Eq.~(\ref{NDEDA1})]$^{39,40}$, we shall
focus on the possibility of achieving such colloidal crystals with
desired SHG signals.


This paper is organized as follows. In section~\ref{form}, we
apply the Ewald-Kornfeld formulation to derive the local electric
field in three typical structures of colloid crystals, and then
perform the NDEDA method to extract the effective linear
dielectric constant and nonlinear susceptibility for SHG. In
Section~\ref{opt}, we discuss the optimization of the
Ewald-Kornfeld summation. Then, we numerically investigate the SHG
under different conditions in Section~\ref{num}, which is followed
by a discussion and conclusion in Section~\ref{dis}.

\section{Formalism}\label{form}

\begin{figure}[h]
\includegraphics[width=200pt]{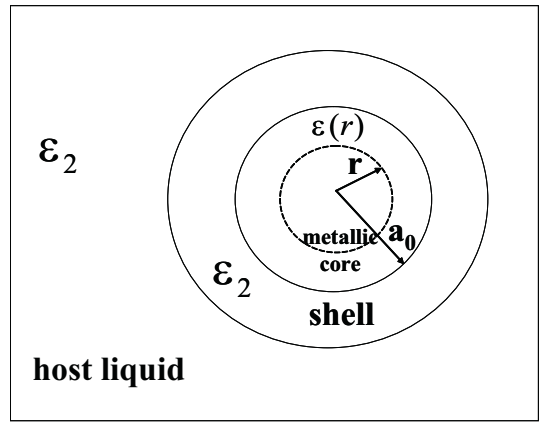}
\caption{/Huang, Jian, Fan,  and Yu}\label{fig1}
\end{figure}

Let us start by considering a graded metallic core with radius
$a_0$ (Fig.~\ref{fig1}). When we take into account quadratic
nonlinearities only, the local constitutive relation between the
displacement field ${\bf D}(r)$ and  electric field ${\bf E}(r)$
 is given
by$^{41-43}$
\begin{equation}
D_i(r) = \sum_j \epsilon_{ij}E_j(r) +
\sum_{jk}\chi_{ijk}(r)E_j(r)E_k(r),\,\, i = x, y, z,
\end{equation}
where $D_i(r)$ and $E_i(r)$ are the $i$th component of ${\bf
D}(r)$ and ${\bf E}(r)$, respectively, and $\chi_{ijk}$ is the
nonlinear susceptibility for SHG. Here $\epsilon_{ij} =
\epsilon(r)\delta_{ij}$ denotes the linear dielectric constant,
which is assumed for simplicity to be isotropic. Both
$\epsilon(r)$ and $\chi_{ijk}(r)$ are functions of $r$, as a
result of the gradation profile along the radius $r$
(Fig.~\ref{fig1}). If a monochromatic external field is applied,
the nonlinearity in the system will generally generate local
potentials and fields at all harmonic frequencies. For a finite
frequency external electric field along $z-$axis of the form
\begin{equation}
E_0 = E_0(\omega) e^{-i\omega t} + c.c.,\label{cc}
\end{equation}
the effective SHG susceptibility $\chi_{2\omega}$ can be extracted
by considering the volume average of the displacement field at the
frequency $2\omega$ in the inhomogeneous medium$^{19,41-43}$. In
Eq.~(\ref{cc}), $c.c.$ is referred to complex conjugate. The
graded metallic core can be built up by adding shells gradually,
making the dielectric constant $\epsilon(r) (r\leq a)$ a radial
function, which is schematically shown in Fig.~\ref{fig1}. We
assume that the dielectric constant of the surface and  the linear
host liquid is a constant for convenience, as mentioned in
Section~\ref{intro}. At radius $r$, the inhomogeneous spherical
particle with $\epsilon(r)$ and $\chi_{2\omega}(r)$ can have the
same dipole moment effect as the homogenous sphere with
$\bar{\epsilon}(r)$ and $\bar{\chi}_{2\omega}(r)$. The equivalent
dielectric constant $\bar{\epsilon}(r)$ can be expressed as the
following differential equation  obtained from the differential
effective dipole approximation method$^{13,39}$,
\begin{equation}
\frac{d\bar{\epsilon}(r)}{dr}=\frac{(\epsilon(r)-\bar{\epsilon}(r))(\bar{\epsilon}(r)+2\epsilon(r))}{r\epsilon(r)}.\label{tartar1}
\end{equation}
On the other hand, the equivalent susceptibility for SHG
$\bar{\chi}_{2\omega}(r)$ can be written as$^{40}$
\begin{eqnarray}
\frac{d\bar{\chi}_{2\omega}(r)}{dr}&=&\bar{\chi}_{2\omega}(r)\left(\frac{2d\bar{\epsilon}_\omega(r)/dr}
{2\epsilon_2+\bar{\epsilon}_\omega(r)}+\frac{d\bar{\epsilon}_{2\omega}(r)/dr}{2\epsilon_2+\bar{\epsilon}_{2\omega}(r)}+\frac{2y(\omega,r)+y(2\omega,r)-3}{r}
\right) \nonumber\\
&&
+3\chi_{2\omega}(r)\left(\frac{[x^2(\omega,r)+\frac{28}{35}z^2(\omega,r)]x(2\omega,r)+\frac{8}{35}[7x(\omega,r)+2z(\omega,r)]z(\omega,r)z(2\omega,r)}{r
f(2\omega,r)f^2(\omega,r)}\right ),\nonumber\\
\label{NDEDA1}
\end{eqnarray}
where
$x(\omega,r)=\frac{\epsilon_2(r)(\bar\epsilon_{\omega}(r)+2\epsilon_{\omega}(r))}{\epsilon_{\omega}(r)(\bar\epsilon_{\omega}(r)+2\epsilon_2(r))},$
$y(\omega,r)=2\frac{(\bar\epsilon_{\omega}(r)-\epsilon_{\omega}(r))(\epsilon_{\omega}(r)-\epsilon_2(r))}{\epsilon_{\omega}(r)(\bar\epsilon_{\omega}(r)+2\epsilon_2(r))},$
$z(\omega,r)=\frac{\epsilon_2(r)(\bar\epsilon_{\omega}(r)-\epsilon_{\omega}(r))}{\epsilon_{\omega}(r)(\bar\epsilon_{\omega}(r)+2\epsilon_2(r))},$
 and
$f(\omega,r)=\frac{3\epsilon_2(r)}{\bar\epsilon_{\omega}(r)+2\epsilon_2(r)}$.
Here the indices $\omega$ and $2\omega$ correspond to basic and
second harmonics for the nonlinear susceptibility, respectively,
and $\bar{\chi}_{2\omega}(r)$ denotes the equivalent SHG
susceptibility of the whole graded spherical particle with radius
$r$.  For convenience, we shall denote $\bar{\chi}_{2\omega}(r=a)$
as $\bar{\chi}_{2\omega}$ in the following.

\begin{figure}[h]
\includegraphics[width=200pt]{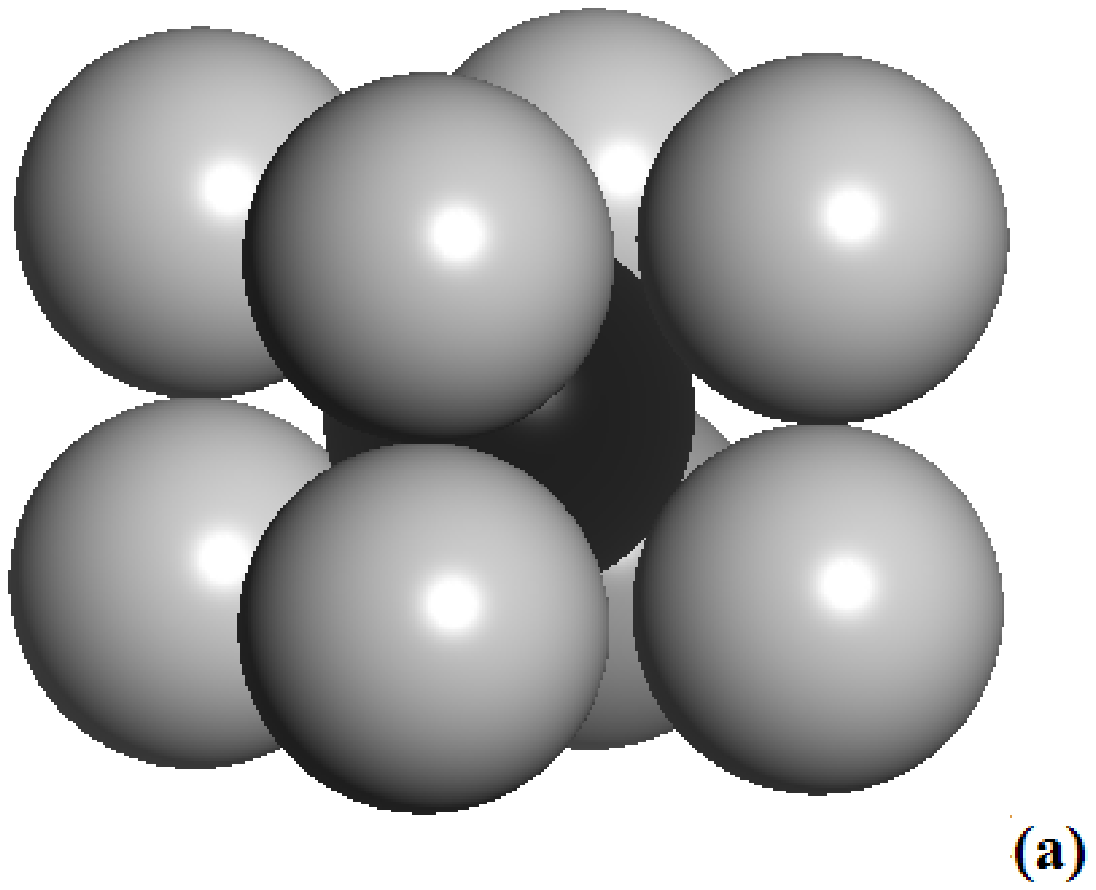}
\includegraphics[width=200pt]{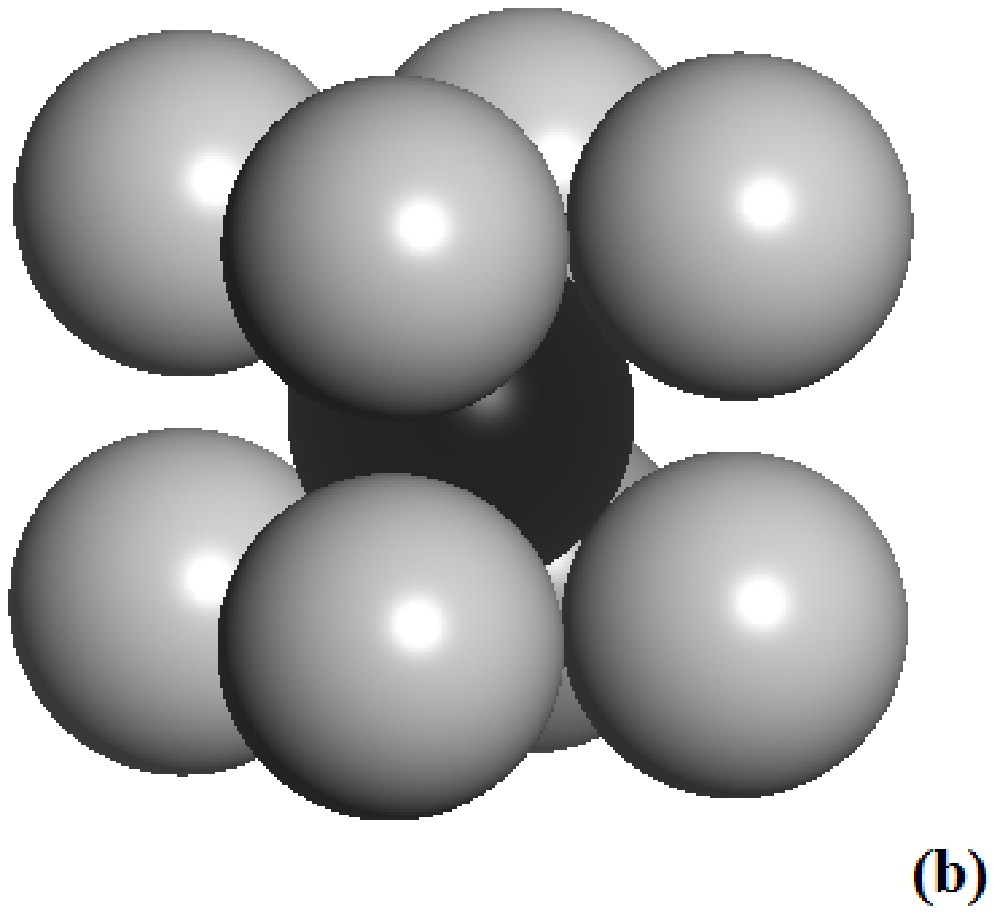}
\includegraphics[width=200pt]{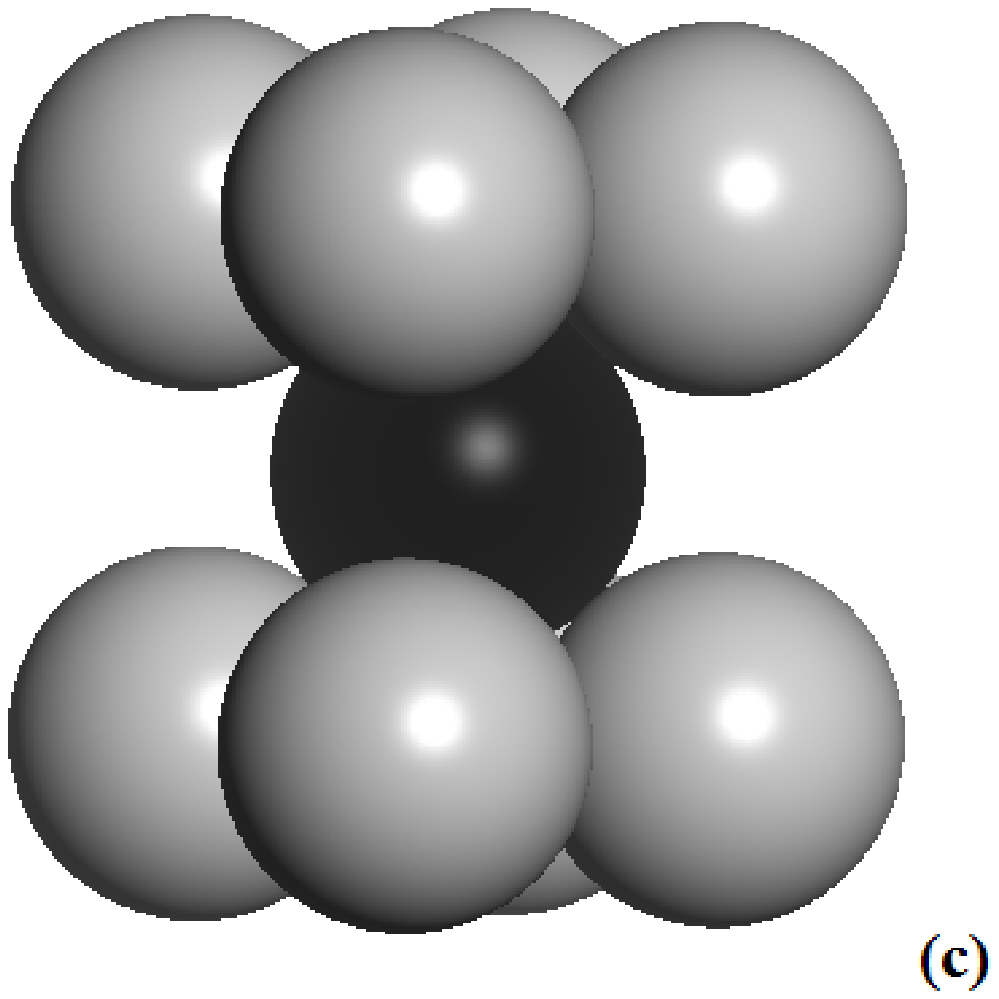}
\caption{/Huang, Jian, Fan,  and Yu}\label{add-fig}
\end{figure}

The above two differential equations
[Eqs.~(\ref{tartar1})-(\ref{NDEDA1})] can be solved numerically as
long as the gradation profiles are given. For obtaining the
effective dielectric constant of the colloidal crystal, we refer
to the Maxwell-Garnett approximation$^{44}$
\begin{equation}
\frac{\epsilon_e-\epsilon_2}{\alpha\epsilon_e+(3-\alpha)\epsilon_2}=\rho
\frac{\bar{\epsilon}-\epsilon_2}{\bar{\epsilon}+2\epsilon_2},
\label{MGA-1}
\end{equation}
where $\rho$ denotes the volume fraction of the metallic component
[see Eq.~(\ref{re1})], and the local field factor $\alpha$
represents $\alpha_{\perp}$ (transverse field cases) and
$\alpha_{\parallel}$ (longitudinal field cases), respectively.
Here the longitudinal (or transverse) field case corresponds to
the fact that the $E$ field of the incident light is parallel (or
perpendicular) to the uniaxial anisotropic axis. For
$\alpha_\perp$ and $\alpha_\parallel$, there is a sum rule
$2\alpha_\perp+\alpha_\parallel=3$ $^{45,46}$. Next, we shall
apply the Ewald-Kornfeld model to compute the local field factor
$\alpha$ for a tetragonal unit cell that can be viewed as a
tetragonal lattice plus a basis of two nanoparticles. One of the
two nanoparticles is located at the corner of the cell, and the
other is at the body center. Without loss of generality, we
consider three representative lattices (Fig.~\ref{add-fig}): the
bct (body-centered tetragonal), bcc (body-centered cubic) and fcc
(face-centered cubic) lattice. If the uniaxial anisotropic axis is
directed along the $z$ axis, the lattice constants can be denoted
by $a (=b)=\ell q^{-\frac{1}{2}}$ along ${\bf x}$ (${\bf y}$) axis
and $c=\ell q$ along ${\bf z}$ axis, and the volume of the unit
cell $V_c = \ell^3.$ The lattice parameters satisfy the geometric
constraint that $a^2+b^2+c^2=16(a_0+d)^2$, when we take into
account  the dielectric surface layer with thickness $d$ on the
graded metallic core. It is easy to obtain the value of $q$ from
their intrinsic structures: $ \textit{q}=0.87358, 1.0$ and
$2^{\frac{1}{3}}$ represent the bct, bcc and fcc lattice,
respectively. Table~\ref{table} shows the calculated values of
$\alpha_{\perp}$ and $\alpha_{\parallel}$ versus $q$, according to
Eq.~(\ref{Ewald1}) below. The degree of anisotropy of the periodic
lattices is measured by how $q$ deviates from unity. Here we
assume that the colloidal particles are packed closely together.
Meanwhile, we obtain a relation between $q$ and the volume
fraction $\rho$ of the metallic component,
\begin{equation}
\rho
=\frac{\pi}{24t^3}\sqrt{\left(\frac{q^3+2}{q}\right)^{3}},\label{re1}
\end{equation}
 with thickness parameter
$t=(a_0+d)/a_0$, $t>1$. The lattice vector of the tetragonal
lattice is given by ${\bf R}=\ell (q^{-\frac{1}{2}}l\hat{{\bf
x}}+q^{-\frac{1}{2}}m\hat{{\bf y}}+qn\hat{{\bf z}}),$ where $l,$
$m,$ and $n$ are integers. When an external electric field
$\mathbf{E_0}$ is applied along the ${\bf x}$ axis, the induced
dipole moment ${\bf p}$ are perpendicular to the uniaxial
anisotropic axis. Considering the field contribution from all the
other particles in the lattice, the local field $\mathbf{E_L}$ at
the lattice point ${\bf r}={\bf 0}$ can be given as
\begin{equation}
E_L = p\sum_{j=1}^2\sum_{{\bf R}\ne {\bf
0}}[-B(R_j)+x_j^2q^2C(R_j)]-\frac{4\pi p}{V_c}\sum_{{\bf G}\neq
{\bf 0}}\Theta({\bf
G})\frac{G_x^2}{G^2}\exp\left(\frac{-G^2}{4\eta^2}\right)+\frac{4p\eta^3}{3\sqrt{\pi}},
\label{Ewald1}
\end{equation}
where $x_j=l-(j-1)/2$, $R_j=|{\bf R}-[(j-1)/2](a\hat{{\bf
x}}+b\hat{{\bf y}}+c\hat{{\bf z}})| ,$ and $\Theta({\bf G}) =
1+\exp [i(u+v+w)/\pi] .$ In Eq.~(\ref{Ewald1}), $B$ and $C$ are
two coefficients given in$^{46}$, $B(r)=\textmd{erfc}(\eta
r)/r^3+2\eta/(\sqrt{\pi}r^2)\exp(-\eta^2 r^2))$ and
$C(r)=3\textmd{erfc}(\eta
r)/r^5+[4\eta^3/(\sqrt{\pi}r^2)+6\eta/(\sqrt{\pi}r^4)]\exp(-\eta^2
r^2)$, where $\textmd{erfc}(\eta r)$ is the complementary error
function and $\eta$ is an adjustable parameter making the
summation converge rapidly. For details, please see
Section~\ref{opt}. In Eq.~(\ref{Ewald1}), $p$ denotes the strength
of the induced dipole moment, and ${\bf G}$  the reciprocal
lattice vector of ${\bf R}$. Thus the local field factor in
transverse fields can be defined as
\begin{equation}
\alpha_{\perp} = \frac{3}{4\pi} \frac{V_cE_L}{p} . \label{alpha0}
\end{equation}
 For the bct, bcc and
fcc lattices, we obtain $\alpha_\bot=0.95351, 1.0$ and $1.0,$
respectively.
Following refs~19 and 20, the effective SHG susceptibility of the
whole system $\chi_{2\omega}$ is given by
\begin{equation}
\chi_{2\omega}=\rho \bar{\chi}_{2\omega} \Gamma(2\omega)
\Gamma^2(\omega),
\end{equation}
where $\Gamma(\omega)$ denotes the factor in a linear system which,
for consistency with Eq.~(\ref{MGA-1}) in getting $\epsilon_e$,
should also be determined by using the Maxwell-Garnett approach.
Thus, we obtain
\begin{equation}
\Gamma(\omega) =
\frac{3\epsilon_2}{(1-\rho\alpha)\bar{\epsilon}(\omega)+(2+\rho\alpha)\epsilon_2}.
\end{equation}
%
Meanwhile, $\bar{\chi}_{2\omega}$ can be obtained through the
NDEDA method [Eq.~(\ref{NDEDA1})].

\section{Optimization of the Ewald-Kornfeld summation
[Eq.~(\ref{Ewald1})]}\label{opt}

In Eq.~(\ref{Ewald1}), we see the adjustable parameter $\eta$
dominates the accuracy and the efficiency of the Ewald-Kornfeld
summation$^{46,47}$. An important aspect of the Ewald-Kornfeld
summation is the tuning in the sense of speed at well controlled
errors. It should be chosen carefully in order to make the
summations both in the real space and the reciprocal lattices
converge rapidly$^{46-48}$. On the other hand, the $r$-space
cutoff $r_{c}$ and the $k$-space cutoff $k_{c}$ are also difficult
to be determined$^{46-48}$. In this section, we shall analyze the
role of these 3 parameters, especially on lattices depicted in
Fig.~\ref{add-fig}. Similarly, it can also be applied to other
lattice models.

In condensed matter physics, complex dielectrics should obey the
famous sum rule $\beta _{x}+\beta _{y}+\beta _{z}=4\pi $, where
$\beta _{x},$ $\beta _{y},$ and $\beta _{z}$ are the local field
factors along $x,$ $y,$ and $z$ axes, respectively
[Eq.~(\ref{beta})]. We shall use this rule to estimate the
accuracy of our algorithm.

First of all, let us investigate how the lattice parameters affect
the accuracy of
calculation. We test 4 lattices with $\{a=1,b=1,c=1,5,10\}$ and $%
\{a=10,b=10,c=10\}$ respectively. The cutoffs in the $r-$ and $k-$
spaces are fixed to 3 which means that we take summation over 6
periods in each direction. By using the relation mentioned above,
we have
\begin{equation}
\beta _{i}=\frac{E_{L,i}({\bf p}_i)V_{c}}{2p_{i}} \equiv
\frac{4\pi}{3} \alpha,\label{beta}
\end{equation}
 where the subscript $i$ stands for Cartesian
directions $x,$ $y,$ and $z.$ The local-field factors $\alpha$
[Eq.~(\ref{MGA-1})] and $\beta$ [Eq.~(\ref{beta})] have exactly
the same concept, the only difference is that there is a
proportionality constant $4\pi/3$ between $\beta$ and $\alpha$, as
shown in [Eq.~(\ref{beta})]. In detail, $\alpha_\perp =(4\pi/3)
\beta_x$ or $(4\pi/3) \beta_y$, and $\alpha_\| = (4\pi/3)
\beta_z$.

\begin{figure}[h]
\includegraphics[angle=270,width=400pt]{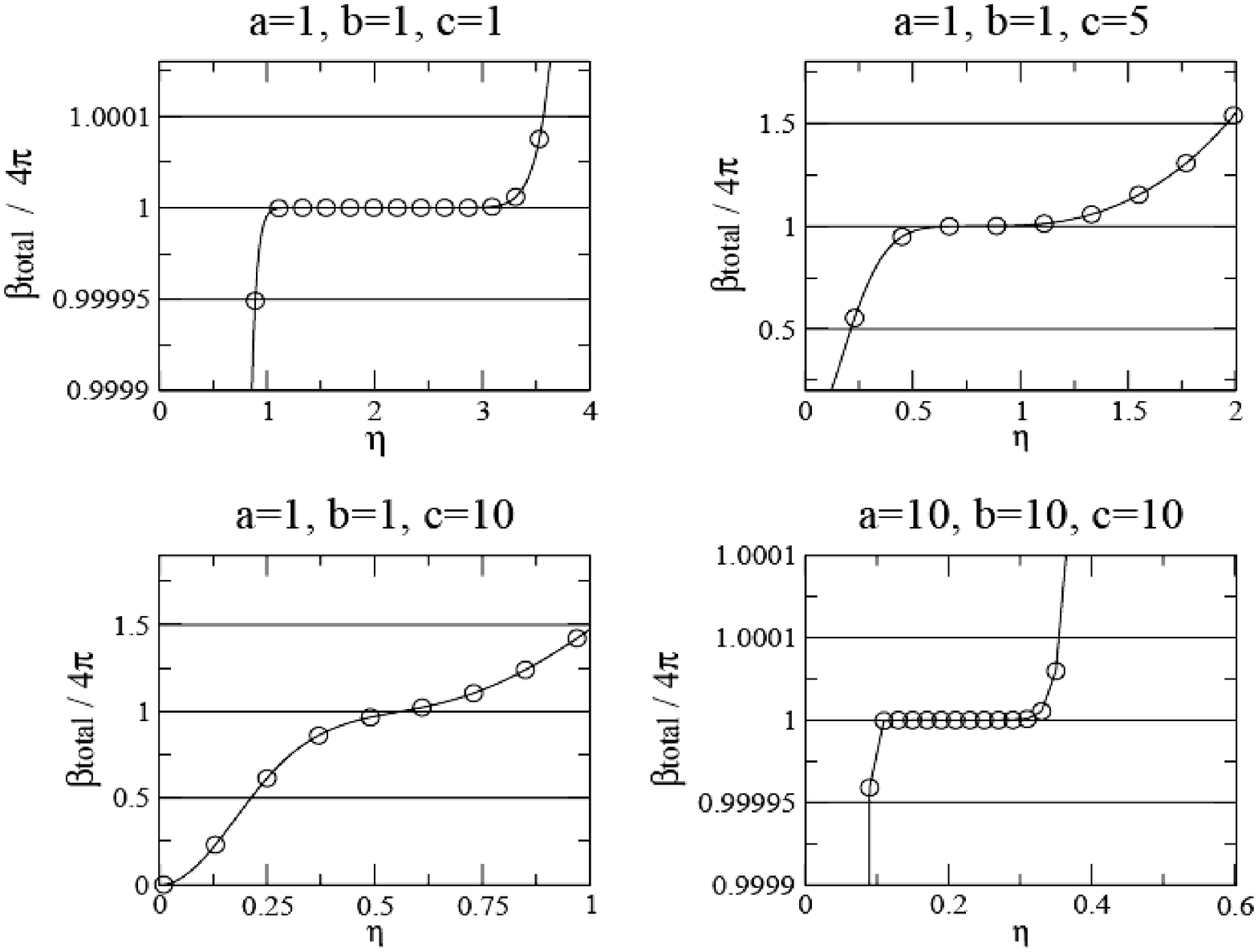}
\caption{/Huang, Jian, Fan,  and Yu}\label{appendix-fig1}
\end{figure}

Next, let us plot the result of the Ewald-Kornfeld summation vs
splitting parameter in Fig.~\ref{appendix-fig1}. The total local
field factor $\beta_{total}=\beta _{x}+\beta _{y}+\beta _{z}$ is
normalized
by $4\pi .$ When $\{{a=b=c=1}\}$, there is a wide platform with $%
\eta $ goes from 1 to 3. In this case, the correctness of our
algorithm can be guaranteed by choosing $\eta $ with any value
within the platform region. However, as ${c}$ increases, the
anisotropic degree becomes strong, and the original platform
rapidly shortens. In the lattice $\{{a=b=1,c=5}\}$, the
proper region of $\eta $ only has the width 0.25. Further, in the lattice $\{{%
a=b=1,c=10}\}$, the platform even disappears. At that time, the
correct result could not be achieved if we would not adjust
$r_{c}$ and $k_{c}$. Another isotropic lattice $\{{a=b=c=10}\}$ is
also investigated.
Although the shape of the figure is similar with that of $\{{%
a=b=c=1}\}$, the platform is remarkably shrunk.

\begin{figure}[h]
\includegraphics[width=400pt]{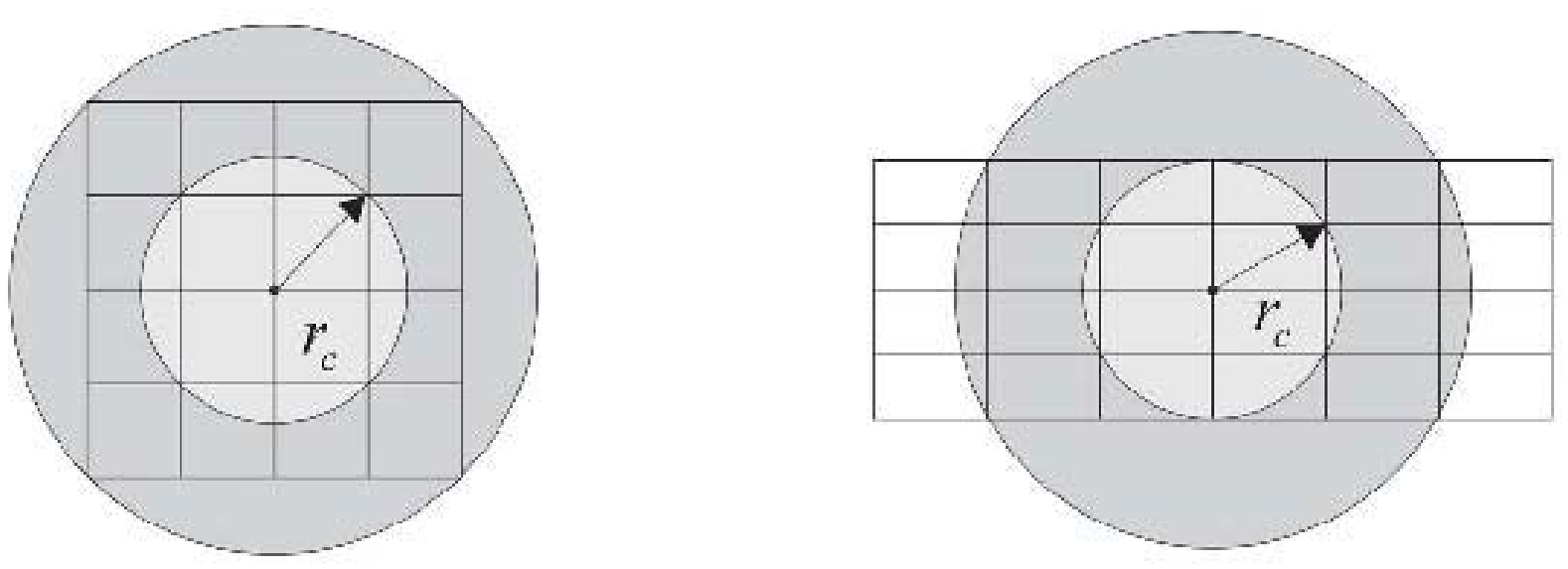}
\caption{/Huang, Jian, Fan,  and Yu}\label{appendix-fig2}
\end{figure}

\begin{figure}[h]
\includegraphics[angle=270,width=400pt]{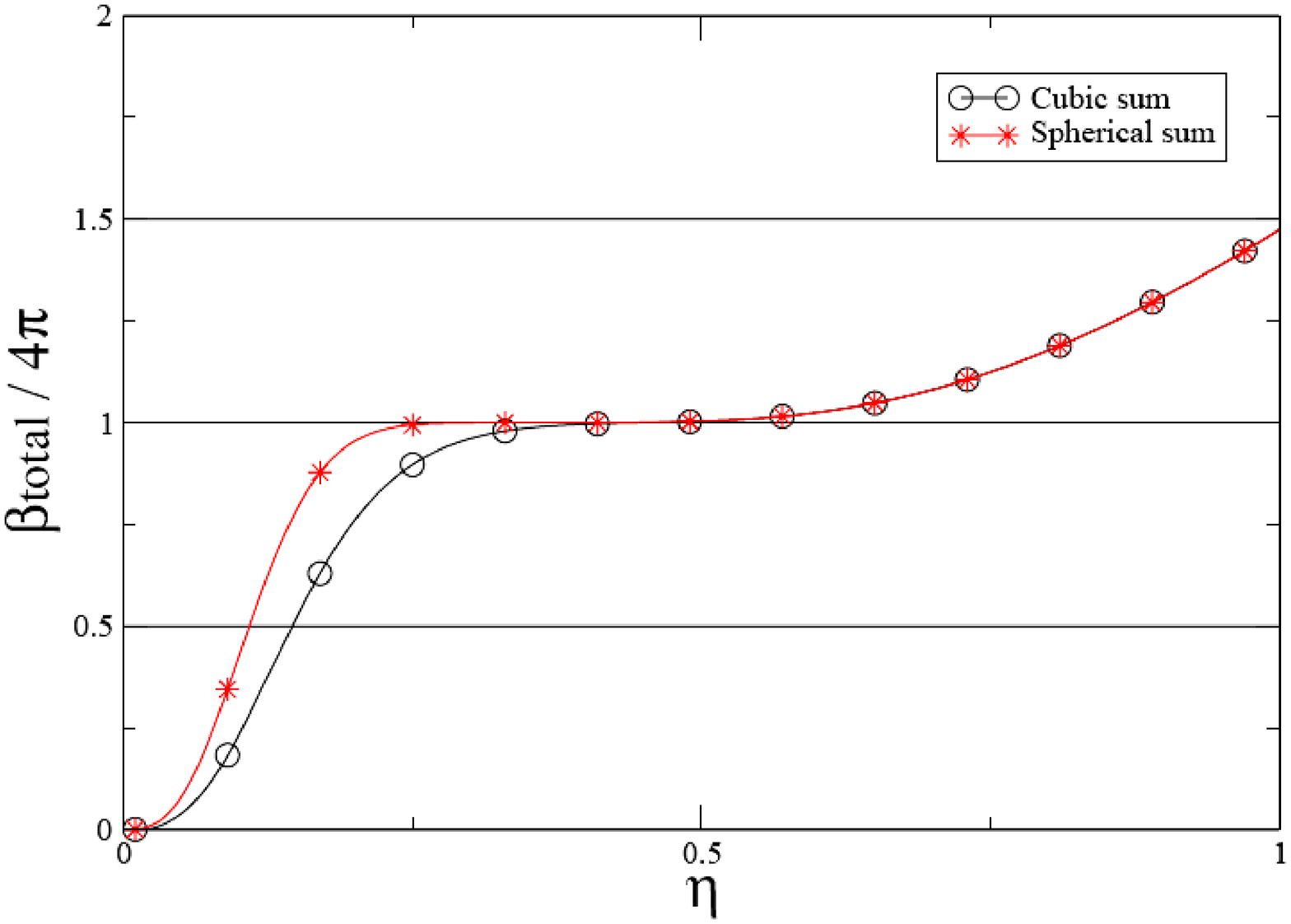}
\caption{/Huang, Jian, Fan,  and Yu}\label{appendix-fig3}
\end{figure}

Then, let us see how the shape of summation region affects the
accuracy. For highly anisotropic lattices, i.e.
$\{{a=b=1,c=10}\}$, the method of cubic region summation is not
applicable because it counts in  many source sites far from the
field point along the uniaxial direction, but ignores many sites
near the field point in the other two isotropic direction (see
Fig.~\ref{appendix-fig2}). It may be improved by using a spherical
region summation. Set $l,$ $m,$ and $n$ to be sum indices in
$r-$space,  and $u,$ $v,$ and $w$ to be  in $k$-space. For the
cubic region summation, we just  require all these indices to be
within [-maximum, maximum]. But for the spherical region
summation, the requirement turns out to be
\begin{equation}
(al)^{2}+(bm)^{2}+(cn)^{2}\leq R_{r}^{2},\label{rr}
\end{equation}%
\begin{equation}
(u/a)^{2}+(v/b)^{2}+(w/c)^{2}\leq R_{k}^{2}.\label{rk}
\end{equation}
Figure~\ref{appendix-fig3} shows the effect of such spherical
region summation. The circle line is obtained by imposing the
restriction that all summation indices are within [-5,5], and the
star line is obtained by using the spherical region summation. The
maximum radii, $R_r$ [Eq.~(\ref{rr})] and $R_k$ [Eq.~(\ref{rk})],
are the length of the diagonal of the unit cell in $r-$ and
$k-$space, respectively,
\begin{equation}
(al)^{2}+(bm)^{2}+(cn)^{2}\leq a^{2}+b^{2}+c^{2} = R_{r}^{2},
\end{equation}%
\begin{equation}
(u/a)^{2}+(v/b)^{2}+(w/c)^{2}\leq (1/a)^{2}+(1/b)^{2}+(1/c)^{2} =
R_{k}^{2}.
\end{equation}%
This summation region is smaller than the cubic region, and less
sites are evaluated naturally. Nevertheless, a wider platform is
achieved for this  region. The main reason is that the spherical
region summation tends to count in the sites which contribute much
to the field point. The two lines in Fig.~\ref{appendix-fig3} show
the same value when $\eta $ becomes larger. Later we shall see
that this is due to the result in the $k$-space summation that is
not modified too much.

\begin{figure}[h]
\includegraphics[angle=270,width=400pt]{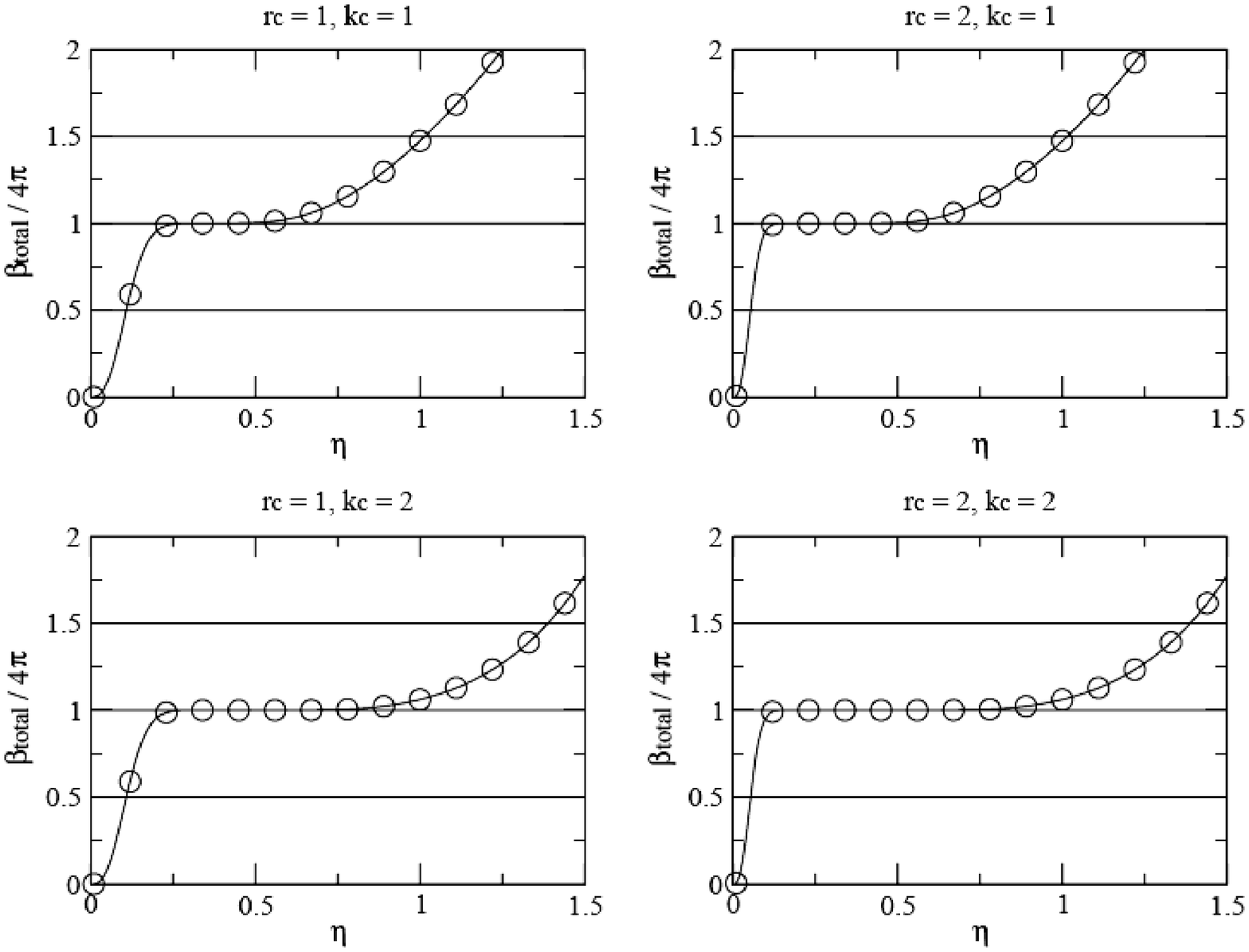}
\caption{/Huang, Jian, Fan,  and Yu}\label{appendix-fig4}
\end{figure}

Last, we demonstrate how $r_{c}$ and $k_{c}$ affect the accuracy.
In general, the larger radii the cutoff has, the more precise the
result is. Unfortunately,  larger summation regions  cost longer
computation time. Thus, we should take optimized values of $r_{c}$
and $k_{c}$. In Fig.~\ref{appendix-fig4}, we calculate the total
local field factor for four configurations of cutoffs. The
left-top graph is copied from Fig.~\ref{appendix-fig3}. Meanwhile,
the right-bottom plot is obtained with twice value of $r_{c}$ and
$k_{c}$. We  find that by increasing $r_{c}$, the platform would
extend to the left with the limit of zero, but there is no such
limit by increasing $k_{c}$. As $k_{c}$ increases, the platform
could extend to the wide space on the right.

In conclusion, in order to optimize the Ewald-Kornfeld summation
[Eq.~(\ref{Ewald1})], we had better perform the sum in a spherical
shape. Enlarging the summation radius in $k$-space is more
efficient than that in $r$-space. Our
method might test the sum rule first with a large enough $r_{c}$ and $%
k_{c}$ (here time is not the main concern), and then it is
convenient for one to choose the center value of $\eta$ in the
platform. This guarantees the correctness of all the computations
performed for achieving Table~\ref{table}.

\section{Numerical Results}\label{num}

For numerical calculations, we set  the linear dielectric constant
of the nonlinear metallic core to have the following Drude form
\begin{equation}
\epsilon(r)=1-\frac{\omega_p(r)^2}{\omega(\omega+i\gamma)},\label{dru}
\end{equation}
where $\omega_p(r)$ means a position-dependent plasma frequency,
and $\gamma$ relaxation rate. For achieving the position-dependent
plasma frequency, one possible way is to fabricate metallic
spherical particles containing multilayers each of which is made
of different metals. For numerical calculations, we take a model
plasma-frequency gradation profile
\begin{equation}\omega_p(r)=\omega_p(0)\left(1-C_\omega
\frac{r}{a}\right),\label{plas}\end{equation} where $C_{\omega}$
is a parameter adjusting the gradation profile.
 For focusing on the
enhancement of the SHG response, we take the intrinsic nonlinear
SHG susceptibility $\chi_{1}$ to be a  frequency-and-position
independent real positive constant.

Figure~\ref{fig2} shows  (a) the imaginary part of the effective
linear dielectric constant (namely, optical absorption), (b) the
real and (c) imaginary parts of the effective SHG susceptibility,
 (d) the modulus of $\bar{\chi}_{2\omega}/\chi_1$, and (e) the
FOM (figure of merit) as a function of frequency for longitudinal
field cases.
 As
$C_\omega$ increases, $\omega_p(r)$ takes on a broader range of
value and leads to a broad plasmon band, while the plasmon peak
shifts to lower frequencies (namely, redshift). The susceptibility
of the SHG also shows an enhancement and the peak of enhancement
can be shifted to lower frequencies, too. Generally, the SHG
susceptibility and FOM can be enhanced in some frequency regions
as $C_{\omega}$ increases.

Figure~\ref{fig3} shows the effective responses and FOM as a
function of thickness parameter $t$, for bct lattices. For the
given lattice it is evident that the effective linear  and
nonlinear optical responses strongly depend on the thickness
parameter. Both the redshift and strength of the plasma resonant
peak  are largest at the smallest $t$ for linear optical
absorption. In this case, the plasma resonant band is also largest
for smallest $t$. Similar behavior can also be found for the
nonlinear SHG responses. On the other hand, $t$ has an effect on
the FOM, too. All of these results come from the combination of
gradation, local fields, and periodic lattice effects. In fact,
the volume fraction for different colloidal lattices also
contributes to the nature of plasma resonant. For the fcc lattice,
its redshift may lie between bct and bcc lattices, and for the
current parameters in use, its deference between fcc and bcc (or
bct) is smaller than 4\% (no pictures shown here).  It can also be
found that as $t$ increases (or, alternatively the colloid
crystals become more dilute), the behavior of plasma resonant
becomes more similar. This can be explained from the dilute limit
approximation model for the enhancement of optical susceptibility.

In Figs.~\ref{fig2}~and~\ref{fig3} the quantities that can be both
positive and negative are plotted in a logarithm of modulus. When
the quantities pass through zero, the logarithm is very large,
thus yielding spikes. In addition, we can reach the conclusion
that the FOM in the high frequency region is still attractive due
to the presence of weak optical absorption.

\section{Discussion and conclusion}\label{dis}


Our main idea is to first reduce the graded metallic cores to
effective ones and then consider colloidal crystals consisting of
such effective nanoparticles embedded in a host liquid that has the
same dielectric constant as the dielectric shells of the
nanoparticles. In doing so , multipolar interaction between the
metallic cores can become unimportant for arbitrary field
polarizations. It should be remarked that, since  the nonlinear
response will depend on the local fields and nonlinear
susceptibility tensors in the whole structure (due to the broken
symmetry at the surface of the metallic core),   the  response can
be tensorial and variant within the core. In this work we have
treated the quantities as a scalar and constant, in order to focus
on the effects of lattices and gradation of our interest.

As the value of $q$ increases, the responses in transverse field
will have slight difference from the longitudinal case (no figures
shown here), because of the little difference between
$\alpha_\bot$ and $\alpha_\|$ for a given structure. From bct, bcc
to fcc lattices, with the increase of $q$ , the longitudinal local
field factor $\alpha_\|$ decreases from 1.09299 at bct lattices to
1.0 at bcc and fcc lattices, while the volume fraction $p$
decreases to  bcc, then increases to fcc. This trend can also be
seen in other bulk samples, such as rhombohedral, orthorhombic and
hexagonal.


We have investigated the cases of graded plasma frequencies, by
assuming the relaxation rate $\gamma$ to be a constant. In fact,
$\gamma$ can also be inhomogeneous. For instance, a
position-dependent profile for  the relaxation rate can  be
achieved experimentally. One possible way may be to fabricate
dirty metallic spherical particles in which the degree of disorder
varies in the radial direction and hence leads to a
relaxation-rate gradation profile. In case of graded relaxation
rates, the nonlinear optical responses can also be adjusted by
choosing appropriate gradation profiles for relaxation
rates$^{11}$.

Throughout the paper, the host medium is assumed to be isotropic.
It is interesting to see what will happen if the host is
anisotropic, e.g., for a graded-index host$^{49}$. In this case,
the gradation is also expected to yield desired enhanced SHG. On
the other hand, optical switching in graded plasmonic crystals via
nonlinear pumping was recently reported$^{50}$, which might also
be realized in graded colloidal crystals proposed in this work.

In summary, based on the Edward-Kornfeld formulation, we have
theoretically exploited a class of nonlinear materials possessing
a nonvalishing SHG susceptibility, which are based on colloidal
crystals of graded metallodielectric nanoparticles. They have been
shown to have an enhancement and redshift of  SHG signals due to
the combination of various effects.


\section*{Acknowledgments}

We thank Professor L. Gao for his assistance in compiling the
computing codes. J.P.H., Y.C.J., and C.Z.F. acknowledge the
financial support by the Shanghai Education Committee and the
Shanghai Education Development Foundation ("Shu Guang" project),
by the Pujiang Talent Project (No.~06PJ14006) of the Shanghai
Science and Technology Committee, by Chinese National Key Basic
Research Special Fund under Grant No. 2006CB921706, and by the
National Natural Science Foundation of China under Grant No.
10604014. K.W.Y. acknowledges financial support through RGC Grant
from the Hong Kong SAR Government.

\clearpage
\newpage


\section*{References and notes}

(1)  Shen, Y. R. {\it The Principles of Nonlinear Optics}; Wiley:
New York, 1984.

(2)  Boyd, R. W. {\it Nonlinear Optics}; Academic: New York, 1992.

(3)  Rodenberger, D. C.;  Heflin, J. R.;  Garito, A. F. {\it
Nature (London)} {\bf 1992}, {\it 359}, 309.

(4)  Fischer, G. L.;  Boyd, R. W.;  Gehr, R. J.;  Jenekhe, S .A.;
Osaheni,  J. A.;  Sipe, J. E.;  Weller-Brophy, L. A. {\it Phys.
Rev. Lett.} {\bf 1995}, {\it 74}, 1871.

(5)  Sekikawa, T.;  Kosuge, A.; Kanai, T.;  Watanabe, S. {\it
Nature (London)} {\bf 2004}, {\it 432}, 605.


(6)  Huang, J. P.; Yu, K. W. {\it Phys. Rep.} {\bf 2006}, {\it
431}, 87.

(7)  Blaaderen, A. V. {\it MRS Bull.} {\bf 2004}, {\it 29}, 85.

(8)  Velikov, K. P.;  Christova, C. G.;  Dullens, R. P. A.; van
Blaaderen, A. {\it Science} {\bf 2002}, {\it 296}, 106.

(9)  Gong T.; Marr, D. W. {\it Appl. Phys. Lett.} {\bf 2004}, {\it
85}, 3760.

(10)  Schilling T.; Frenkel, D. {\it Phys. Rev. Lett.} {\bf 2004},
{\it 92}, 085505.

(11)  Huang J. P.; Yu, K. W. {\it Appl. Phys. Lett.} {\bf 2004},
{\it 85}, 94.

(12) Bennink, R. S.;  Yoon, Y.-K.;  Boyd, R. W.; Sipe, J. E. {\it
Opt. Lett.} {\bf 1999}, {\it 24}, 1416.


(13)  Milton, G. W. {\it The Theory of Composites}; Cambridge
University Press: Cambridge, 2002; Chap.~VII.

(14) Fan, C. Z.; Huang, J. P.; Yu, K. W. {\it J. Phys. Chem. B}
{\bf 2006}, {\it 110}, 25665.

(15) Wei, E. B.;  Song, J. B.; Gu, G. Q. {\it J. Appl. Phys.} {\bf
2004}, {\it 95}, 1377.

(16)  Sang Z. F.; Li, Z. Y. {\it Opt. Commun.} {\bf 2006}, {\it
259}, 174.

(17) Wei, E. B.; Dong, L.; Yu, K. W. {\it J. Appl. Phys.} {\bf
2006}, {\it 99}, 054101.

(18) Gu, G. Q.; Yu, K. W. {\it J. Appl. Phys.} {\bf 2003}, {\it
94}, 3376.

(19)  Hui P. M.; Stroud, D. {\it J. Appl. Phys.} {\bf 1997} {\it
82}, 4740.

(20)  Fan C. Z.; Huang, J. P. {\it Appl. Phys. Lett.} {\bf 2006},
{\bf 89}, 141906.

(21) Reis, H. {\it J. Chem. Phys.} {\bf 2006}, {\it 125}, 014506.

(22) Nappa, J.; Russier-Antoine, I.;  Benichou, E.;  Jonin, C.;
Brevet, P. F. {\it  J. Chem. Phys.} {\bf 2006}, {\it 125}, 184712.

(23)  Shalaev, V. M. {\it  Phys. Rep.} {\bf 1996}, {\it 272}, 61.

(24)  Chiang, H. P.;  Leung, P. T.;  Tse, W. S. {\it J. Phys.
Chem. B} {\bf 2000}, {\it 104}, 2348.

(25)  Han, J.;  Chen, D.;  Ding, S.;  Zhou, H.; Han, Y.;  Xiong,
G.; Wang, Q. {\it J. Appl. Phys.} {\bf 2006} {\it 99}, 023526.

(26)  Pezzetta, D.; Sibilia, C.;   Bertolotti, M.;  Ramponi, R.;
Osellame, R.;  Marangoni, M.;  Haus, J. W.; Scalora, M.; Bloemer,
M. J.;  Bowden, C. M. {\it J. Opt. Soc. Am. B} {\bf 2002}, {\it
19}, 2102.

(27) Purvinis, G.;   Priambodo, P. S.;  Pomerantz, M.;  Zhou, M.;
Maldonado, T. A.;
 Magnusson, R. {\it Opt. Lett.} {\it 29}, 1108.

(28)  Dadap, J. I.;  Shan, J.;  Eisenthal, K. B.; Heinz, T. F.
{\it Phys. Rev. Lett.} {\bf 1999} {\it 83}, 4045.

(29)  Yang, N.; Angerer, W. E.;  Yodh, A. G. {\it Phys. Rev.
Lett.} {\bf 2001}, {\it 87}, 103902.

(30)  Jen, S. H.;  Dai, H. L. {\it J. Phys. Chem. B} {\bf 2006},
{\it 110}, 23000.

(31)  Son, D. H.;  Wittenberg, J. S.;  Banin, U.; Alivisatos, A.
P. {\it J. Phys. Chem. B} {\bf 2006}, {\it 110}, 19884.

(32)  Xu, P.;  Ji, S. H.;  Zhu, S. N.;  Yu, X. Q.; Sun, J.;
Wang, H. T.;  He, J. L.;  Zhu, Y. Y.; Ming, N. B. {\it Phys. Rev.
Lett.} {\bf 2004}, {\it 93}, 133904.

(33)  Bernal R.; Maytorena, J. A. {\it Phys. Rev. B} {\bf 2004},
{\it 70}, 125420.

(34) {\it Colloids and Colloid Assemblies}; edited by Caruso, F.;
Wiley-VCH: Weinheim, 2004.

(35)  Nehl, C. L.;  Grady, N. K.;  Goodrich, G. P.; Tam, F.;
Halas, N. J.;  Hafner, J. H. {\it Nano Lett.} {\bf 2004}, {\it 4},
2355.

(36)  Mitzi, D. B.;  Kosbar, L. L.;  Murray, C. E.; Copel, M.;
Afzali, A. {\it Nature (London)} {\bf 2004}, {\it 428}, 299.

(37)  Huang, J. P.; Yu, K. W. {\it Appl. Phy. Lett.} {\bf 2005},
{\it 87}, 071103.


(38)  Ewald, P. P. {\it Ann. Phys. (Leipzig)} {\bf 1921}, {\it
64}, 253; Kornfeld, H. {\it Z. Phys.} {\bf 1924}, {\it 22}, 27.

(39)  Gao, L.;  Huang, J. P.; Yu, K. W. {\it Phys. Rev. B} {\bf
2004}, {\it 69}, 075105, and references therein.

(40)  Gao, L.; Yu, K. W. {\it Phys. Rev. B} {\bf 2005}, {\it 72},
075111. Although this reference presents all components for an
effective nonlinear susceptibility of SHG, in our work, for
simplicity, we only perform numerical calculations on the $z$
component, thus yielding eq 4.

(41)  Hui, P. M.;  Xu, C.;  Stroud, D. {\it Phys. Rev. B} {\bf
2004}, {\it 69}, 014202.

(42)  Hui, P. M.;  Xu, C.;  Stroud, D. {\it Phys. Rev. B} {\bf
2004}, {\it 69}, 014203.

(43) Huang, J. P.;  Hui, P. M.;  Yu, K. W. {\it Phys. Lett. A}
{\bf 2005}, {\it 342}, 484.

(44)  Lo, C. K.;  Wan, J. T. K.; Yu, K. W. {\it J. Phys.: Condens.
Matter} {\bf 2001}, {\it 13}, 1315.


(45)  Landau, L. D.;  Lifshitz, E. M.; Pitaevskii, L. P. {\it
Electrodynamics of Continuous Media}; 2nd ed.; Pergamon:  New
York, 1984; Chap.~II.

(46)  Lo, C. K.; Yu, K. W. {\it Phys. Rev. E} {\bf 2001},  {\it
64}, 031501.

(47)  Shen, L. {\it Dynamic electrorheological effects of rotating
spheres}; MPhil thesis; Chinese University of Hong Kong: Hong
Kong, 2005.

(48) Wang,  Z.; Holm, C. {\it J. Chem. Phys.} {\bf 2001}, {\it
115}, 6351.

(49)  Xiao, J. J.; Yu, K. W. {\it Appl. Phys. Lett.} {\bf 2006},
{\it 88}, 071911.

(50) Xiao, J. J.;  Yakubo, K.; Yu, K. W. {\it Appl. Phys. Lett.}
{\bf 2006}, {\it 88}, 241111.


\clearpage
\newpage

\begin{table}
\caption{Values of $\alpha_{\perp}$ and $\alpha_{\parallel}$
computed at different $q$. $q = 0.87358,$ $1,$ and $2^{1/3}$
correspond to bct, bcc, and fcc lattices, respectively. }
\begin{tabular}{llll}
\hline\hline
q & $\alpha_{\perp}$ & $\alpha_{\parallel}$ \\
\hline
0.87358 & 0.953506 & 1.09299\\
 0.9    & 0.971231&1.05754& \\
 1.0    & 1& 1& \\
 1.1    & 0.999345& 1.00131& \\
 1.2    &0.996275 & 1.00745& \\
$2^{1/3}$    & 1&1 & \\
 1.3    & 1.00601& 0.987988& \\
 1.4    &1.03492 & 0.930155& \\
 1.5    & 1.08376&0.832478 & \\
 1.6    & 1.15032 & 0.699352& \\
 1.7    &1.23137 & 0.537268& \\
 1.8    & 1.32368&0.352638 & \\
 1.9    & 1.42459& 0.150817 & \\
 \hline\hline
\end{tabular}
\label{table}
\end{table}

\clearpage
\newpage
\section*{Figure captions}

Fig.~1. Schematic graph showing the graded metallic core with
radius $a_0$ embedded in a linear host liquid. The metallic core
has a dielectric shell that has the same dielectric constant as
the host liquid, and the core can be built up by adding shells
gradually. $\epsilon_2$ denotes the linear dielectric constant of
the host liquid and shell, and  $\epsilon(r)$ is the
radius-dependent dielectric constant of the graded metallic core.

Fig.~2. Schematic graph showing unit cells of (a) bct, (b) bcc,
and (c) fcc lattices with lattice constants $a$, $b$, and $c$,
which satisfy $a (=b) =\ell q^{-1/2}$ and $c = \ell q$. Here,
$q=0.87358$ (bct), $1.0$ (bcc), and $2^{1/3}$ (fcc).

Fig.~3. Normalized total local field factors versus splitting
parameters in different lattices. The $\eta$ corresponding to
$\beta_{{\rm total}}/(4\pi) =1$ leads to accurate result of the
Ewald-Kornfeld summation [Eq.~(\ref{Ewald1})].

Fig.~4. Cubic and spherical summation region for lattices with
different anisotropic degree.

Fig.~5. Normalized total local field factors versus splitting
parameters in cubic and spherical region summation. The platform
for the $\eta$ corresponding to $\beta_{{\rm total}}/(4\pi) =1$ is
widened by using spherical region summation.

Fig.~6. Normalized total local field factors versus splitting
parameters for various $r_c$ and $k_c$ cutoffs. Increasing $r_c$
or $k_c$ yields the left or right extension of the platform of the
$\eta$ that corresponds to $\beta_{{\rm total}}/(4\pi) =1$,
respectively.

Fig.~7. For the bct lattice (longitudinal field), (a) the linear
optical absorption Im[$\epsilon_e(\omega)$], (b)
Im[$\chi_{2\omega}/\chi_1$], (c) Re[$\chi_{2\omega}/\chi_1]$, (d)
modulus of $\chi_{2\omega}/\chi_1$, and (e) the
FOM=$|\chi_{2\omega}|/[\chi_1{\rm Im}(\epsilon_e)]$ versus the
normalized incident angular frequency of $\omega/\omega_p(0)$ for
the dielectric function gradation profile [Eq.~(\ref{dru})] with
various plasma-frequency gradation profiles [Eq.~(\ref{plas})]:
$C_\omega=0.3,$ $0.5,$ and $0.7.$  Here $|\cdots|$ denotes the
absolute value or modulus of $\cdots$. Parameters:
$\gamma=0.02\omega_p(0),$ $t=3,$ and $\epsilon_2=2.25$.

\begin{figure}[h]
\includegraphics[width=200pt]{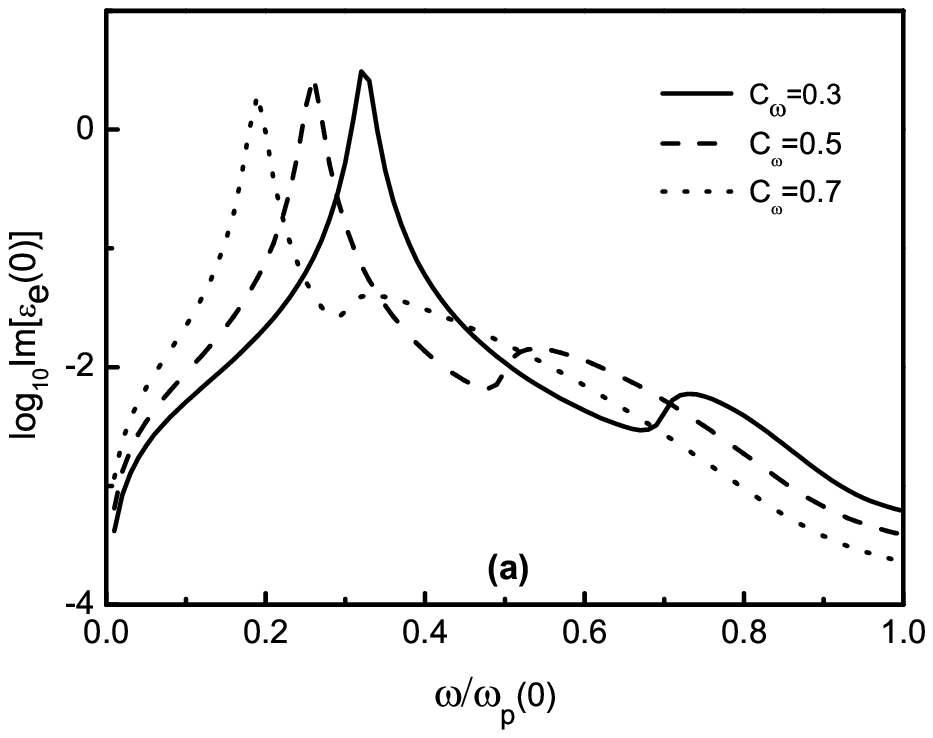}
\includegraphics[width=200pt]{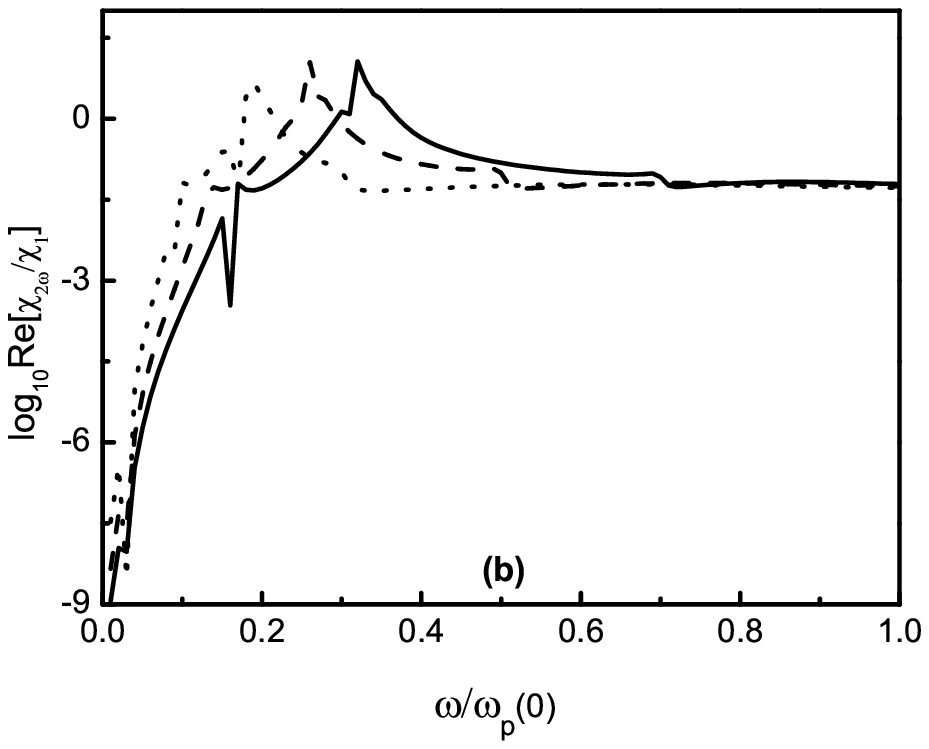}\\
\includegraphics[width=200pt]{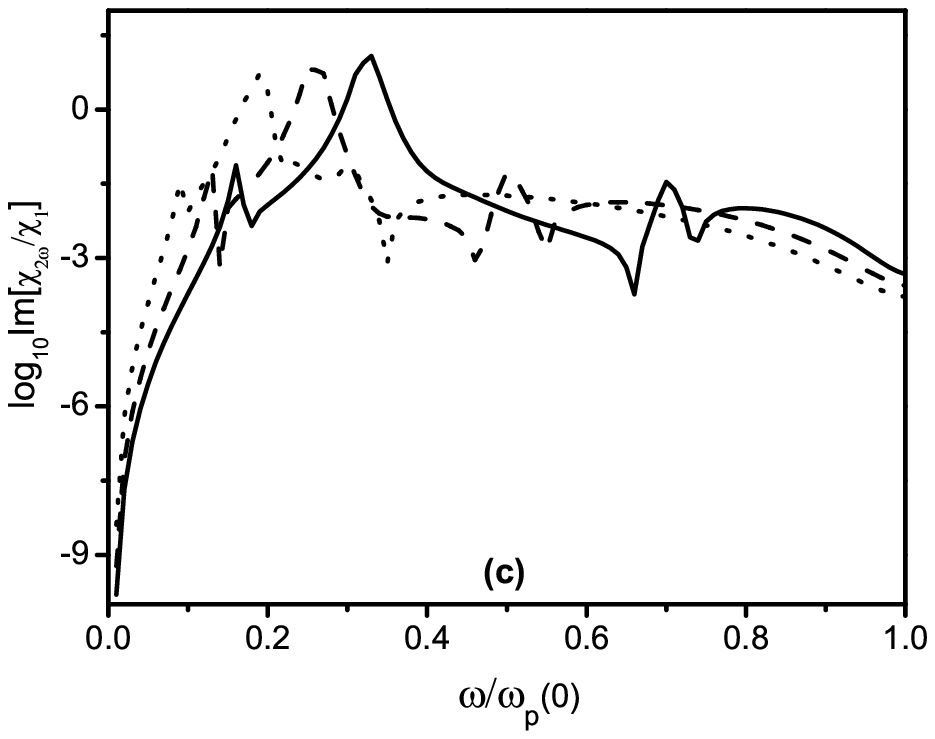}
\includegraphics[width=200pt]{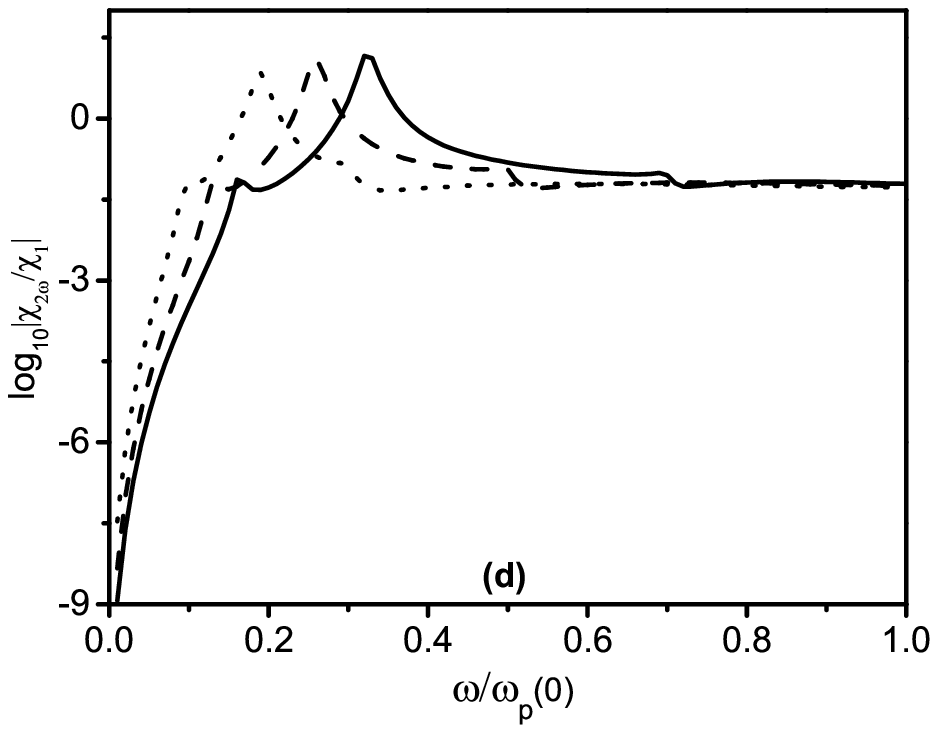}\\
\includegraphics[width=200pt]{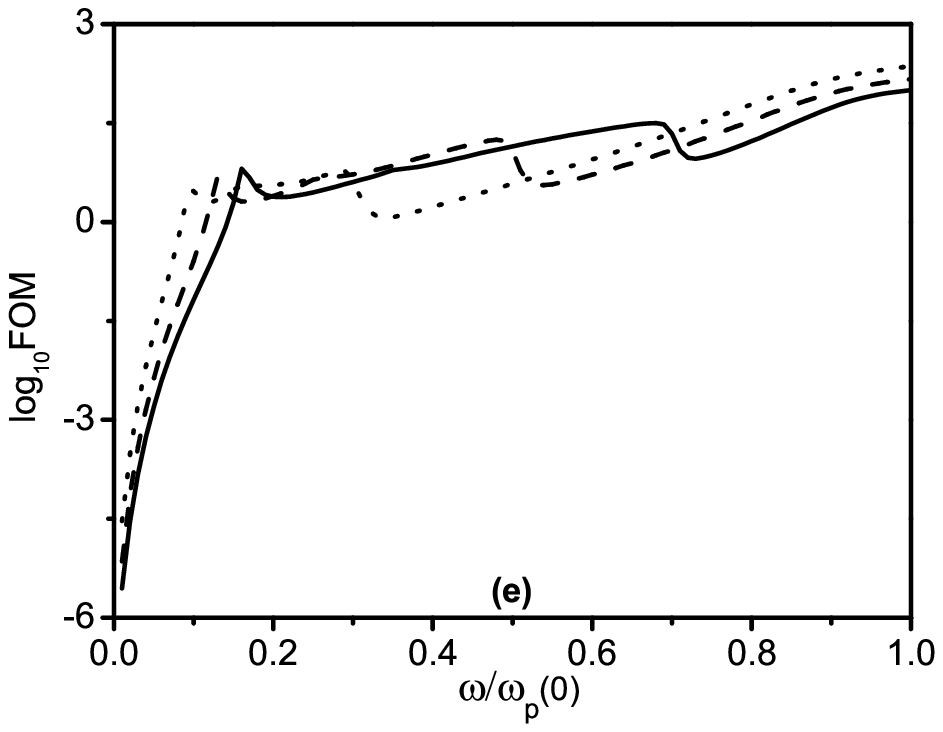}
\caption{/Huang, Jian, Fan, and Yu}\label{fig2}
\end{figure}

\begin{figure}[h]
\includegraphics[width=200pt]{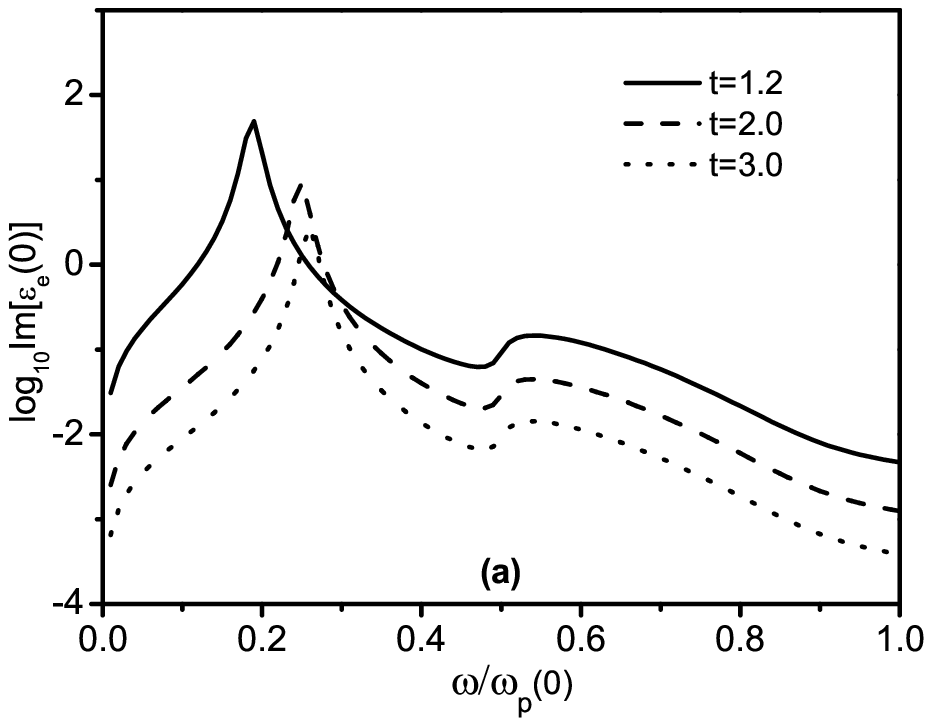}
\includegraphics[width=200pt]{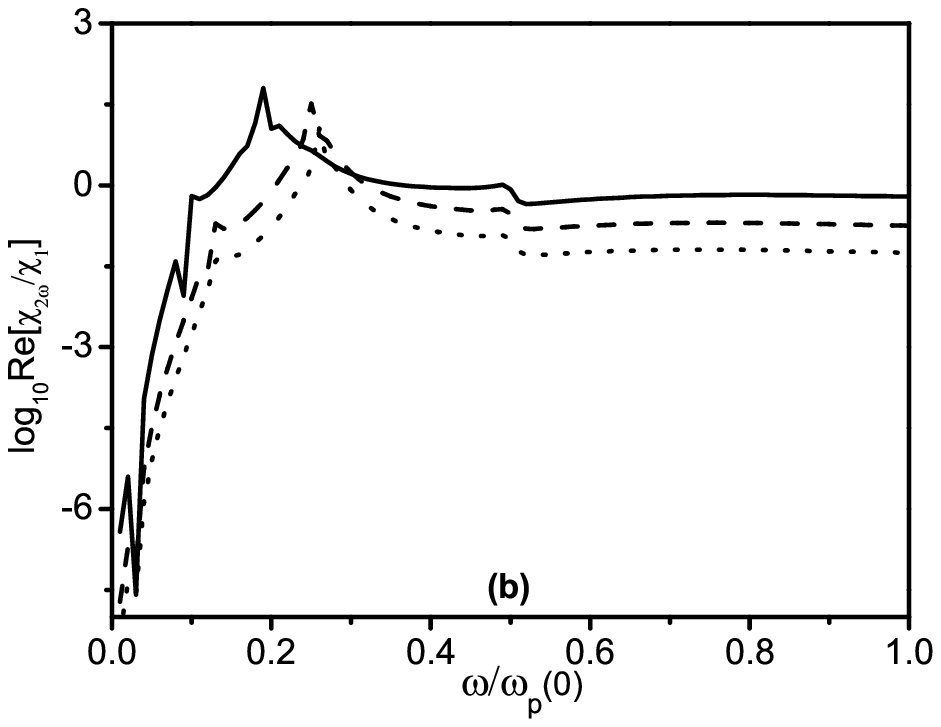}
\includegraphics[width=200pt]{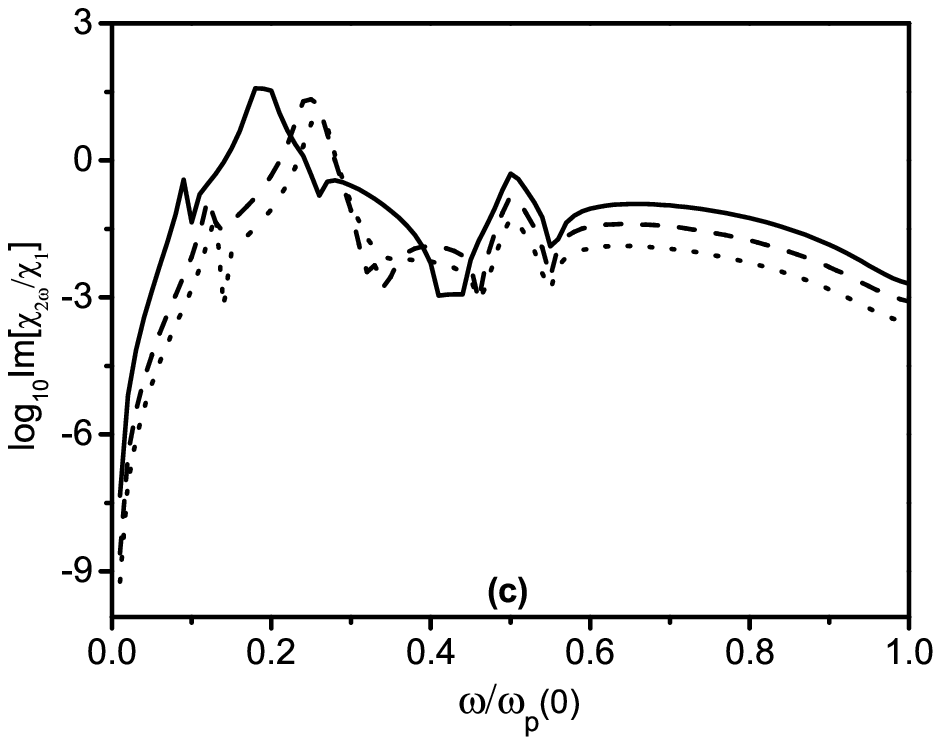}
\includegraphics[width=200pt]{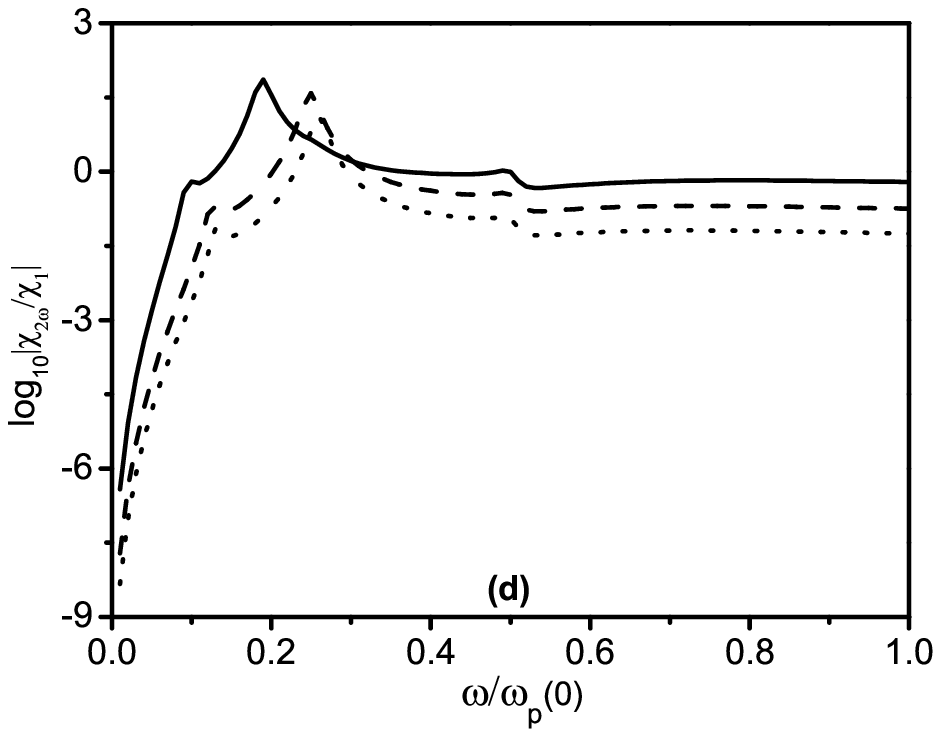}
\includegraphics[width=200pt]{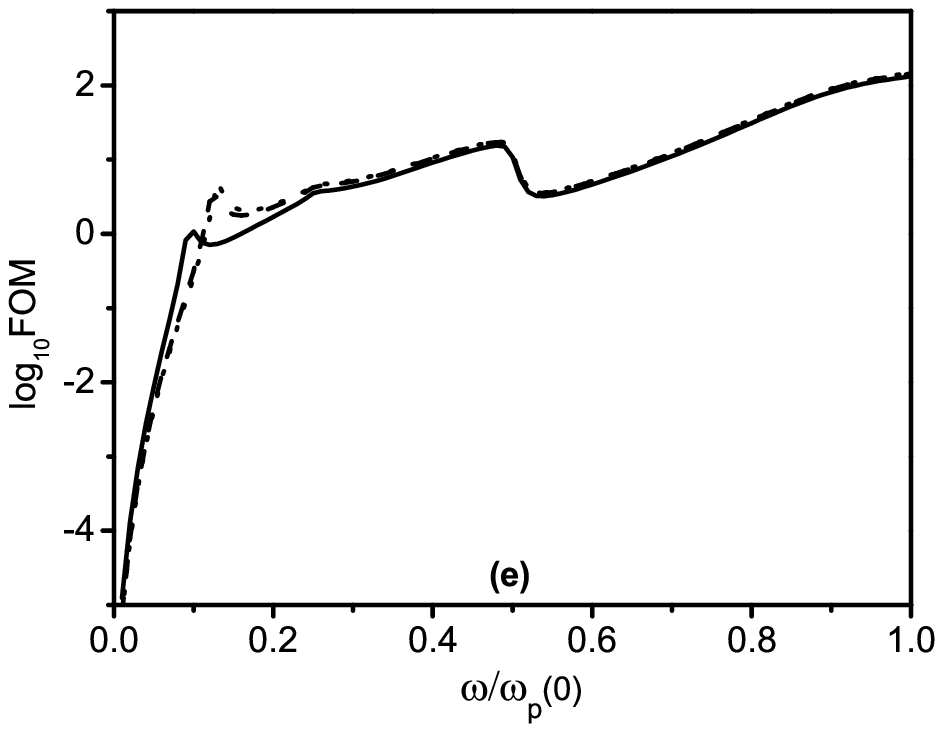}
\caption{/Huang, Jian, Fan,  and Yu}\label{fig3}
\end{figure}

Fig.~8. Same as Fig.~\ref{fig2}, but for different thickness
$t=1.2,$ $2.0,$ and $3.0.$  Parameters: $\gamma=0.02\omega_p(0)$,
$C_{\omega}=0.5,$ and $\epsilon_2=2.25$.

\clearpage
\newpage

\newpage

\newpage

\newpage

\newpage

\newpage

\newpage

\newpage

\end{document}